# DEUTERIUM MICROBOMB ROCKET PROPULSION [1]

F. Winterberg

University of Nevada, Reno

[1] Presented in part at the NASA-JPL-AFRL 2008 Advanced Space Propulsion Workshop, Oct. 28-30, Pasadena, California.

Abstract

Large scale manned space flight within the solar system is still confronted with the solution of two problems: 1. A propulsion system to transport large payloads with short transit times between different planetary orbits. 2. A cost effective lifting of large payloads into earth orbit. For the solution of the first problem a deuterium fusion bomb propulsion system is proposed where a thermonuclear detonation wave is ignited in a small cylindrical assembly of deuterium with a gigavolt-multimegampere proton beam, drawn from the magnetically insulated spacecraft acting in the ultrahigh vacuum of space as a gigavolt capacitor. For the solution of the second problem, the ignition is done by argon ion lasers driven by high explosives, with the lasers destroyed in the fusion explosion and becoming part of the exhaust.

## 1. Introduction

Since 1954 I have been actively working on inertial confinement fusion at a time this research was still classified in the US. I had independently discovered the basic principles and had presented them in 1955 at a meeting organized by von Weizsaecker of the Max Planck Institute in Goettingen. The abstracts of this meeting still exist and are kept in the library of the University of Stuttgart. My paper dealt with the problem of the non-fusion ignition of thermonuclear reactions by convergent shockwaves and imploding shells, including the application for rocket propulsion.

With chemical propulsion manned space flight to the moon is barely possible and only with massive multistage rockets. For manned space flight beyond the moon, nuclear propulsion is indispensible. Nuclear thermal propulsion is really not much better than advanced chemical propulsion. Ion propulsion, using a nuclear reactor driving an electric generator, has a much higher specific impulse, but not enough thrust for short interplanetary transit times, as they are needed for manned missions. This leaves the propulsion by a chain fission bombs (or fission triggered fusion bomb) as the only credible option. There the thrust and specific impulse are huge in comparison. But a comparatively small explosive yield is there be desirable. Making the yield too small, the bombs become extravagant in the sense that only a small fraction of the fission explosive is consumed. The way to overcome this problem is the non-fission ignition of small fusion explosions. A first step in this direction is the non-fission ignition of deuterium-tritium (DT) thermonuclear micro-explosions; the easiest one to be ignited expected to be realized in the near future. Because of it, I had chosen this reaction for the first proposed thermonuclear micro-explosion propulsion concept, with the ignition was done by an intense relativistic electron beam [1, 2]. But because in the DT reaction 80% of the energy is released into neutrons which cannot be reflected from the spacecraft by a magnetic mirror, it was proposed to surround the micro-explosion with a neutron-absorbing hydrogen propellant, increasing the thrust on the expense of the specific impulse. It was for this reason that in the "Daedalus" interstellar probe study of the British Interplanetary Society [3], the neutron-less helium3-deuterium ($He^3$-D) reaction was proposed, because for such a mission the specific impulse should be as high as possible. But even in a $He^3$-D plasma, there are a some neutron producing DD reactions. But there is no large source of $He^3$ on the earth, even though it might exist on the surface of the moon, and the atmosphere of Jupiter. Much less energy goes into neutrons, but it is more difficult to ignite the DD reaction. The situation is illustrated in Fig 1. On the left side it shows the experimentally verified ignition of a DT pellet with the X-rays drawn in an underground test from a fission bomb (Centurion Halite experiment at the Nevada Test Site). It required for the ignition of a few megajoule, with a density$\times$radius target product, $\rho r > 1\text{g/cm}^2$. And on the right side is the 15 Megaton "Mike" test, where with the Teller-Ulam configuration a large ball of liquid deuterium is ignited by a fission bomb. For the DD reaction one has $\rho r \geq 10\text{g/cm}^2$. In between is the proposed hypothetical deuterium target, where a detonation wave in a thin cylindrical deuterium rod is ignited by a pulsed $10^7$ Ampere-GeV proton beam, utilizing the strong magnetic field of the beam current.

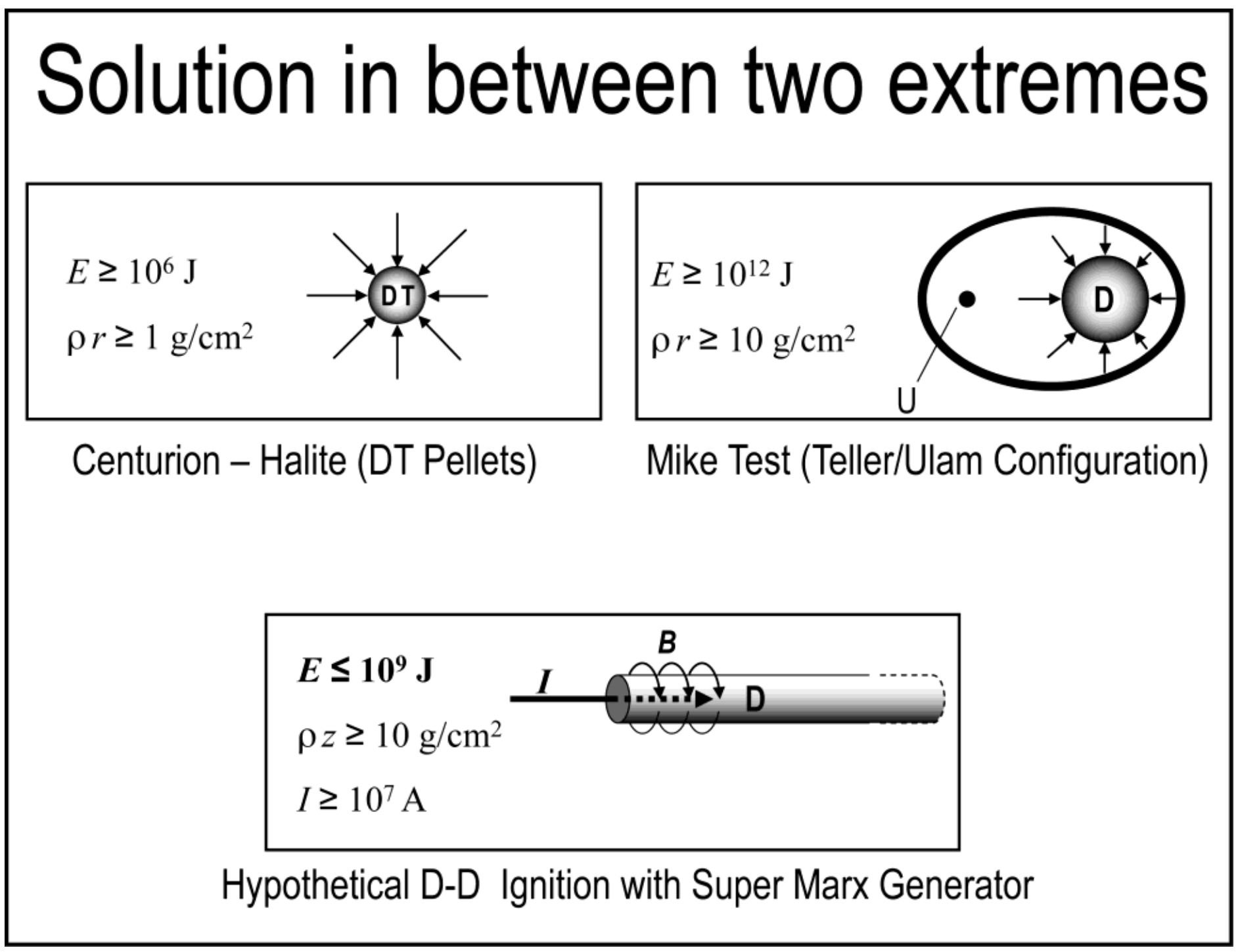

Fig. 1 GeV – Megampere proton beam.

To estimate the order of magnitude what is needed, we consider a spacecraft with a mass of $M_o = 10^3\,\text{tons} = 10^9\,\text{g}$, to be accelerated by one $g \cong 10^3\,\text{cm/s}^2$, with a thrust $T = M_o g \cong 10^{12}$ dyn. To establish the magnitude and number of fusion explosions needed to propel the spacecraft to a velocity of $v = 100$ km/s = $10^7$ cm/s, we use the rocket equation

$$T = c\frac{dm}{dt} \tag{1}$$

where we set $c \cong 10^8$ cm/s, equal to the expansion velocity of the fusion bomb plasma.
We thus have

$$\frac{dm}{dt} = \frac{T}{c} = 10^4\ g/s = 10\,kg/s \tag{2}$$

The propulsion power is given by

$$P = \frac{c^2}{2}\frac{dm}{dt} = \frac{c}{2}T \tag{3}$$

in our example it is $P = 5\times10^{19}$ erg/s.

With $E = 4\times10^{16}$ erg equivalent to the explosive energy of one ton of TNT, $P$ is equivalent to about one nuclear kiloton bomb per second.

From the integrated rocket equation

$$v = c\ln\left(1+\frac{\Delta M}{M_o}\right) \tag{4}$$

where $v$ is the velocity reached by the spacecraft after having used up all the bombs of mass $\Delta M$, one has for $\Delta M << M_o$,

$$\Delta M/M_o \cong v/c \tag{5}$$

If one bomb explodes per second, its mass according to (2) is $m_o = 10^4$ g.

Assuming that the spacecraft reaches a velocity of $v = 100$ km/s $= 10^7$ cm/s, the velocity needed for fast interplanetary travel, one has $\Delta M = 10^8$ g, requiring $N = \Delta M/m_o = 10^4$ one kiloton fusion bombs, releasing the energy $E_b = 5\times10^{19}\times10^4 = 5\times10^{23}$ erg. By comparison, the kinetic energy of the spacecraft is $E_s = (1/2)M_o v^2 = 5\times10^{23}$ erg, about 10 times less. In reality it is still smaller, because a large fraction of the energy released by the bomb explosions is dissipated into space.

One can summarize these estimates by concluding that a very large number of nuclear explosions is needed, which for fission explosions, but also for deuterium-tritium explosions, would become very expensive. This strongly favors deuterium, more difficult to ignite in comparison to a mixture of deuterium with tritium, but abundantly available. Here I will try to show how bomb propulsion solely with deuterium might be possible.

## 2. The Non-Fission Ignition of Small Deuterium Nuclear Explosives

With no deuterium-tritium (DT) micro-explosions yet ignited, the non-fission ignition of pure deuterium (D) fusion explosions seems to be a tall order. An indirect way to reach this goal is by staging a smaller DT explosion with a larger D explosion. There the driver energy, but not the driver may be rather small. A direct way requires a driver with order of magnitude larger energies.

I claim that the generation of GeV potential wells, made possible with magnetic insulation of conductors levitated in ultrahigh vacuum, has the potential to lead to order of magnitude larger driver energies [4, 5]. It is the ultrahigh vacuum of space by which this can be achieved. And if the spacecraft acting as a capacitor is charged up to GeV potentials, there is no need for its levitation.

If charged to a positive GeV potential, a gigajoule intense relativistic ion beam below the Alfven current limit can be released from the spacecraft and directed on the D explosive for its ignition. If the current needed for ignition is below the Alfvén limit for ions, the beam is "stiff". The critical Alfvén current for protons is $I_A = 3.1 \times 10^7 \beta\gamma [\mathrm{A}]$ , for GeV protons well in excess of the critical current to entrap the DD fusion reaction products, a condition for detonation [6].

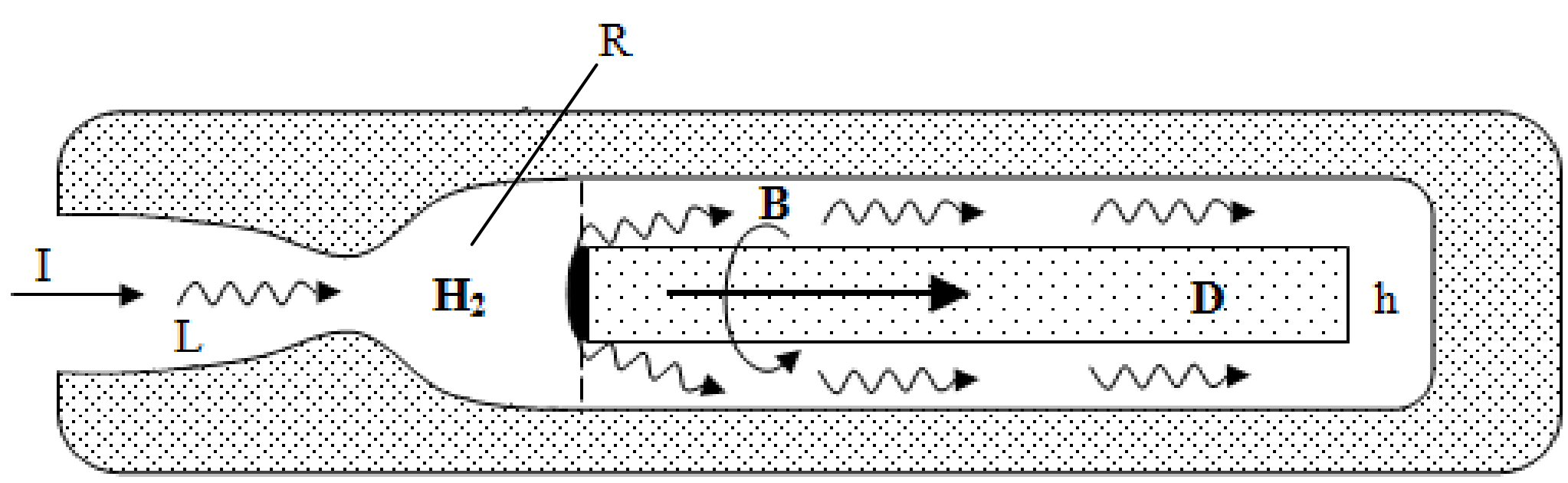

**Figure 2:** Pure deuterium fusion explosion ignited with an intense ion beam. **D** deuterium rod, **h** hohlraum, **I** ion beam, **B** magnetic field, **R** miniature target rocket chamber, $\mathbf{H_2}$ solid hydrogen, **L** laser beam to heat hydrogen in miniature rocket chamber.

In a possible bomb configuration shown in Fig.2, the liquid (or solid) D explosive has the shape of a long cylinder, placed inside a cylindrical "hohlraum" **h**. A GeV proton beam **I** coming from the left, in entering the hohlraum dissipates part of its energy into a burst of X-rays compressing and igniting the D bomb-cylinder. With its gigajoule energy lasting less than $10^{-7}$ s, the beam power is greater than $10^{16}$ Watt, sufficiently large to ignite the D explosive. The main portion of the beam energy is focused by the cone onto the deuterium rod, igniting at its end a detonation wave. The beam is stopped over a short distance by the proton-deuterium two stream instability [7], enhanced by a collision-less magneto-hydrodynamic shock [8].

In DT the condition for propagating burn in a spherical target of radius r and density ρ, is given by

$$\rho r \geq 1 \text{g/cm}^2 \tag{6}$$

whereas for the D-D reaction the condition is

$$\rho r \geq 10 \text{ g/cm}^2 \tag{7}$$

With the optimal ignition temperature of D-D reaction about 10 times larger than for the DT reaction, and assuming the same density for the compressed deuterium as for deuterium-tritium, it follows that the energy for ignition of a deuterium sphere with a radius 10 times larger than for a deuterium-tritium sphere is $10^4$ larger, that is from a few megajoule for a DT sphere to about $10^4$ megajoule.[2] Obviously, no laser or particle beam can easily reach these kinds of energies. The situation is changed in a fundamental way for a thin deuterium rod of length $z$, ignited by an intense ion beam with a current of ~ $10^7$ Ampere entrapping the charged fusion reaction products. There the condition (7) is replaced by

$$\rho z \geq 10 \text{ g/cm}^2 \tag{8}$$

Even if the density is less than ~ 100 g/cm$^3$, corresponding to $10^3$-fold compressed liquid hydrogen, the smaller target density can be made up easily by a sufficiently long deuterium cylinder.

The stopping range of the GeV protons by the two stream instability alone is given by

$$\lambda \cong \frac{1.4c}{\varepsilon^{1/3}\omega_i} \tag{9}$$

where $c$ is here is the velocity of light, and $\omega_i$ the proton ion plasma frequency, furthermore $\varepsilon = n_b/n$, where $n_b$ is the proton number density in the proton beam, and $n$ the deuterium target number density. If the cross section of the beam is 0.1 cm$^2$, one obtains for a $10^7$ Ampere beam that $n_b = 2\times10^{16}$ cm$^{-3}$. For a 100-fold compressed deuterium rod one has $n = 5\times10^{24}$ cm$^{-3}$ with $\omega_i = 2\times10^{15}$ s$^{-1}$. One there finds that $\varepsilon = 4\times10^{-9}$ and $\lambda \cong 1.2\times10^{-2}$ cm. This short length, together with the formation of a collision-less magneto-hydrodynamic shock, ensures the dissipation of the beam energy into a small volume at the end of the deuterium rod. At a deuterium number density $n = 5\times10^{24}$ cm$^{-3}$, one has $\rho = 17$ g/cm$^3$, and to have $\rho z \geq 10$ g/cm$^2$, thus requires that $z \geq 0.6$ cm. With $\lambda < z$, the condition for the ignition of a thermonuclear detonation wave is satisfied. With $T \approx 10^9$ K, the ignition energy is given by

$$E_{ign} \sim 3nkT\pi r^2 z \tag{10}$$

[2] With a convergent shock wave ignition in the center of the compressed deuterium sphere this energy is less, but even then still much more than a few megajoule.

For 100-fold compressed deuterium, one has $\pi r^2 = 10^{-2}$ cm$^2$, when initially it was $\pi r^2 = 10^{-1}$ cm$^2$. With $\pi r^2 = 10^{-2}$ cm$^{-2}$, $z$ = 0.6 cm, one finds that $E_{ign} \leq 10^{16}$ erg or ≤ 1 gigajoule. This energy is provided by the $10^7$ Ampere - Gigavolt proton beam lasting $10^{-7}$ seconds. The time is short enough to assure the cold compression of deuterium to high densities. For a $10^3$-fold compression, found feasible in laser fusion experiments, the ignition energy is ten times less.

## 3. Delivery of a GeV Proton Beam onto the Deuterium Fusion Explosive

The spacecraft is positively charged against an electron cloud surrounding the craft, and with a magnetic field of the order $10^4$ G, easily reached by superconducting currents flowing in an azimuthal direction, it is insulated against the electron cloud up to GeV potentials. The spacecraft and its surrounding electron cloud form a virtual diode with a GeV potential difference. To generate a proton beam, it is proposed to attach a miniature hydrogen filled rocket chamber **R** to the deuterium bomb target, at the position where the proton beam hits the fusion explosive (see Fig.2). A pulsed laser beam from the spacecraft is shot into the rocket chamber, vaporizing the hydrogen, which is emitted through the Laval nozzle as a supersonic plasma jet. If the nozzle is directed towards the spacecraft, a conducting bridge is established, rich in protons between the spacecraft and the fusion explosive. Protons in this bridge are then accelerated to GeV energies, hitting the deuterium explosive. Because of the large dimension of the spacecraft, the jet has to be aimed at the spacecraft not very accurately.

The original idea for the electrostatic energy storage on a magnetically insulated conductor was to charge up to GeV potentials a levitated superconducting ring, with the ring magnetically insulated against breakdown by the magnetic field of a large toroidal current flowing through the ring. It is here proposed to give the spacecraft a topologically equivalent shape, using the entire spacecraft for the electrostatic energy storage (see Fig.3). There, toroidal currents flowing azimuthally around the outer shell of the spacecraft, not only magnetically insulate the spacecraft against the surrounding electron cloud, but the currents also generate a magnetic mirror field which can reflect the plasma of the exploding fusion bomb. In addition, the expanding bomb plasma can induce large currents, and if these currents are directed to flow through magnetic field coils positioned on the upper side of the spacecraft, electrons from there can be emitted into space surrounding the spacecraft by thermionic emitters placed on the inner side of these coils in a process called inductive charging [9]. It recharges the spacecraft for subsequent proton beam ignition pulses. A small high voltage generator driven by a small onboard fission reactor can make the initial charging, ejecting from the spacecraft negatively charged pellets.

With the magnetic insulation criterion, $B > E$, ( $B, E$, in electrostatic units) where $B$ is the magnetic field surrounding the spacecraft, then for $B \cong 10^4$G, $E = 3\times10^3$ esu = $9\times10^5$ V/cm, one has $E \sim (1/3)B$. A spacecraft with the dimension $l \sim 3\times10^3$ cm, can then be charged to a potential $El \sim 3\times10^9$ Volts. The stored electrostatic energy is of the order $\varepsilon \sim \left(E^2/8\pi\right) l^3$.

For $E = 3\times10^3$ esu, and $l = 3\times10^3$ cm, it is of the order of one gigajoule. The discharge time is of the order $\tau \sim l/c$, where $c = 3\times10^{10}$ cm/s is the velocity of light. In our example we have $\tau \sim 3\times10^{-8}$ sec. For a proton energy pulse of one gigajoule, the beam power is $3\times10^{16}$ erg/s = 30 petawatt, large enough to ignite a pure deuterium explosion.

## 4. Lifting of Large Payloads into Earth Orbit

To lift large payloads from the surface of the earth into orbit remains the most difficult task. For it a different approach is proposed.

For a launch from the surface of the earth magnetic insulation inside the earth atmosphere fails, and with it the proposed pure D bomb configuration. Here a different technique is suggested, I had first proposed in a classified report, dated January of 1970 [10], declassified July 2007. A similar idea was proposed in a classified Los Alamos report, dated November 1970, [11], declassified July 1979. The idea is to use a replaceable laser for the ignition of each nuclear explosion, with the laser material thereafter becoming part of the propellant. The Los Alamos scientists had proposed to use for this purpose an infrared carbon dioxide ($CO_2$) or chemical laser, but this idea does not work, because the wavelength is there too long, and therefore unsuitable for inertial confinement fusion. I had suggested an ultraviolet argon ion laser instead. However, since argon ion lasers driven by an electric discharge have a small efficiency, I had suggested a quite different way for its pumping, illustrated in Fig.4, where the efficiency can be expected to be quite high. As shown in Fig.4, it was proposed to use a cylinder of solid argon, surrounding it by a thick cylindrical shell of high explosive. If simultaneously detonated from outside, a convergent cylindrical shockwave is launched into the argon. For the high explosive one may choose hexogen with a detonation velocity of 8 km/s. In a convergent cylindrical shockwave the temperature rises as $r^{-0.4}$, where $r$ is the distance from axis of the cylindrical argon rod. If the shock is launched from a distance of ~1 m onto an argon rod with a radius equal to 10 cm, the temperature reaches 90,000 K, just right to excite the upper laser level of argon. Following its heating to 90,000 K the argon cylinder radially expands and cools, with the upper laser level frozen into the argon. This is similar as in a gas dynamic laser, where the upper laser level is frozen in the gas during its isentropic expansion in a Laval nozzle. To reduce depopulation of the upper laser level during the expansion by super-radiance, one may dope to the argon with a saturable absorber, acting as an "antiknock" additive. In this way megajoule laser pulses can be released within 10 nanoseconds. A laser pulse from a small Q-switched argon ion laser placed in the spacecraft can then launch a photon avalanche in the argon rod, igniting a DT micro-explosion.

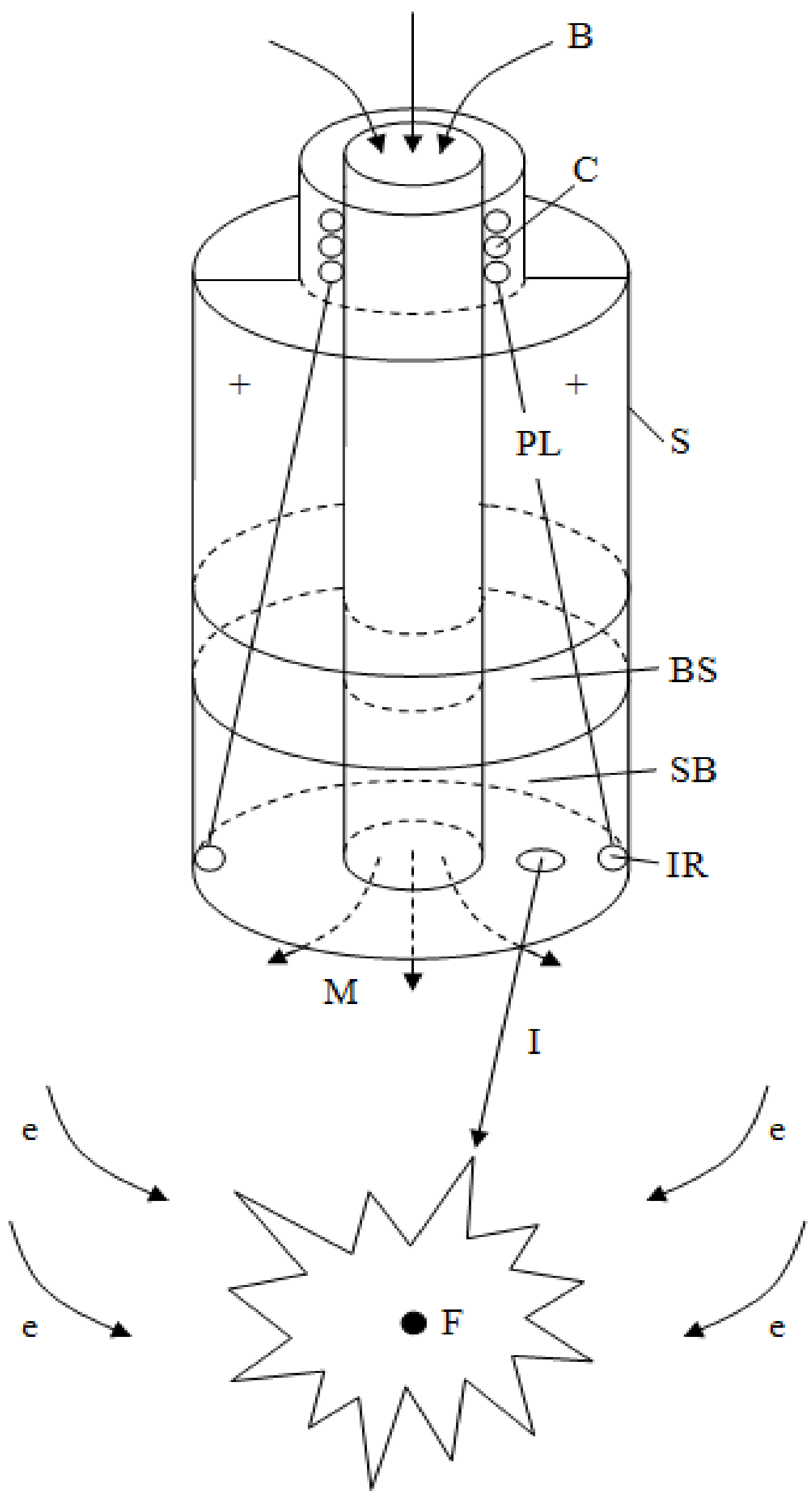

**Figure 3:** Superconducting "atomic" spaceship, positively charged to GeV potential, with azimuthal currents and magnetic mirror **M** by magnetic field **B**. **F** fusion minibomb in position to be ignited by intense ion beam **I**, **SB** storage space for the bombs, **BS** bioshield for the paylod **PL**, **C** coils pulsed by current drawn from induction ring **IR**. **e** electron flow neutralizing space charge of the fusion explosion plasma.

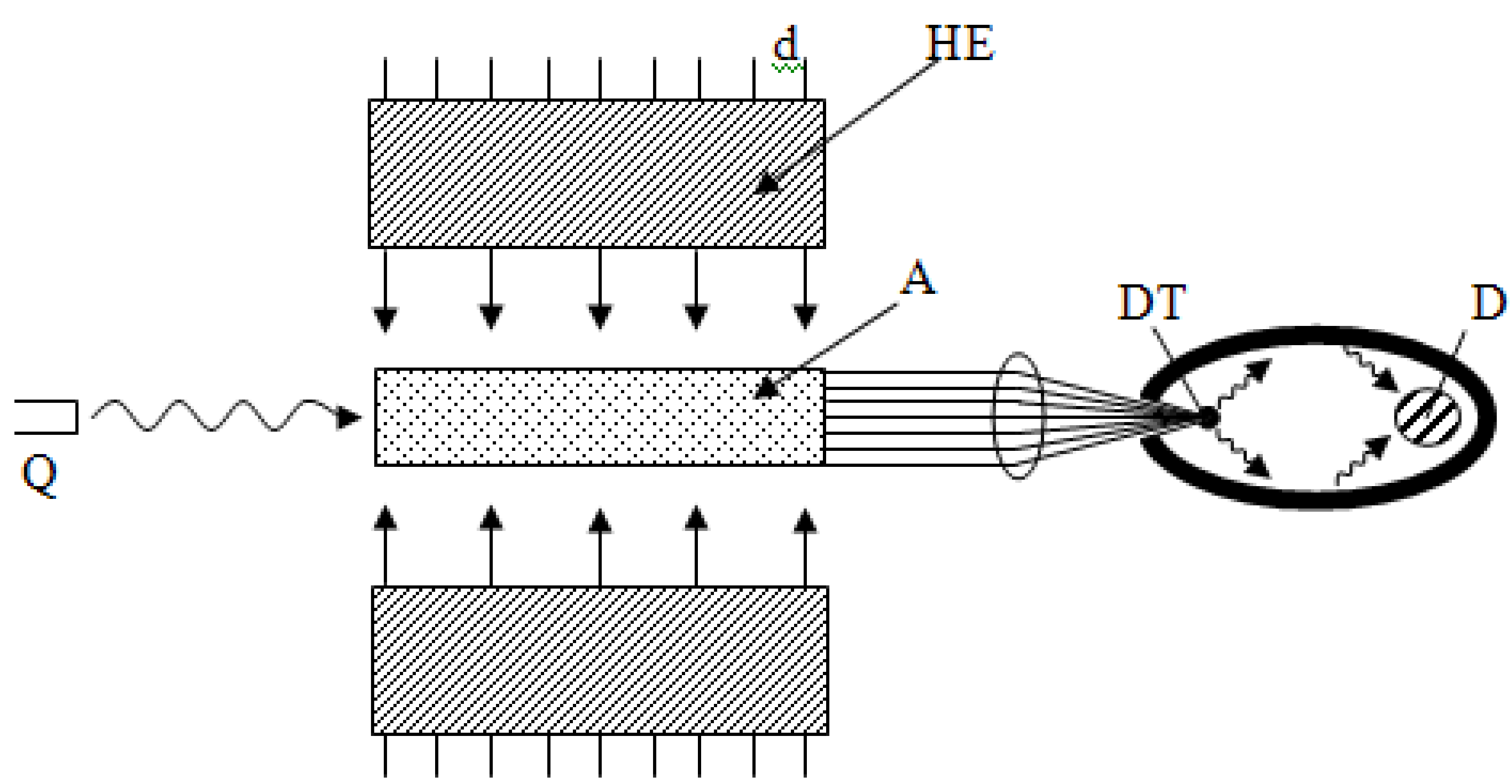

Fig.4: Argon ion laser igniter, to ignite a staged **DT →D** fusion explosion in a mini-Teller-Ulam configuration. **A** solid argon rod. **HE** cylindrical shell of high explosive, **d** detonators. **Q** Q-switched argon ion laser oscillator.

Employing the Teller-Ulam configuration, by replacing the fission explosive with a DT micro-explosion, one can then ignite a much larger D explosion.

As an alternative one may generate a high current linear pinch discharge with a high explosive driven magnetic flux compression generator. If the current $I$ is of the order $I = 10^7$ [A], the laser can ignite a DT thermonuclear detonation wave propagating down the high current discharge channel, which in turn can ignite a much larger pure D explosion.

If launched from the surface of the earth, one has to take into account the mass of the air entrained in the fireball. The situation resembles a hot gas driven gun, albeit one of rather poor efficiency. There the velocity gained by the craft with $N$ explosions, each setting off the energy $E_b$, is given by

$$v = \sqrt{2N E_b / M_o} \qquad (11)$$

For $E_b = 5\times10^{19}\,\mathrm{erg}, M_o = 10^9\,\mathrm{g}$, and setting for $v$ = 10 km/s = $10^6$ cm/s the escape velocity from the earth, one finds that $N \geq 10$. Assuming an efficiency of 10%, about 100 kiloton explosions would there be needed.

**<u>References</u>**